\documentclass[aip,apl,numerical,reprint]{revtex4-1}

\usepackage[normalem]{ulem}
\usepackage{graphicx}
\usepackage{graphics}
\usepackage{bm}
\usepackage{epstopdf}

\begin{document}

\title{Reconfigurable Gradient Index using VO$_2$ Memory Metamaterials} 

\author{M. D. Goldflam}
\email[]{mgoldfla@physics.ucsd.edu}
\affiliation{Physics Department, University of California, San Diego, La Jolla, CA 92093, USA}
\author{T. Driscoll}
\affiliation{Center for Metamaterials and Integrated Plasmonics, Pratt School of Engineering, Duke University. Durham, NC, 27708, USA}
\affiliation{Physics Department, University of California, San Diego, La Jolla, CA 92093, USA}
\author{B. Chapler}
\author{O. Khatib}
\affiliation{Physics Department, University of California, San Diego, La Jolla, CA 92093, USA}
\author{N. Marie Jokerst}
\author{S. Palit}
\author{D. R. Smith}
\affiliation{Center for Metamaterials and Integrated Plasmonics, Pratt School of Engineering, Duke University. Durham, NC, 27708, USA}
\author{Bong-Jun Kim}
\affiliation{Metal-Insulator Transition Creative Research Lab, ETRI, Daejeon 305-350, Republic of Korea}
\author{Giwan Seo}
\affiliation{School of Advanced Device Technology, University of Science and Technology, Daejeon 305-350, Republic of Korea}
\author{Hyun-Tak Kim}
\affiliation{Metal-Insulator Transition Creative Research Lab, ETRI, Daejeon 305-350, Republic of Korea}
\affiliation{School of Advanced Device Technology, University of Science and Technology, Daejeon 305-350, Republic of Korea}
\author{M. Di Ventra}
\author{D. N. Basov}
\affiliation{Physics Department, University of California, San Diego, La Jolla, CA 92093, USA}
\date{\today}

\begin{abstract}
We demonstrate tuning of a metamaterial device that incorporates a form of spatial gradient control.  Electrical tuning of the metamaterial is achieved through a vanadium dioxide layer which interacts with an array of split ring resonators. We achieved a spatial gradient in the magnitude of permittivity, writeable using a single transient electrical pulse. This induced gradient in our device is observed on spatial scales on the order of one wavelength at 1 THz. Thus we show the viability of elements for use in future devices with potential applications in beamforming and communications.
\end{abstract}

\maketitle 

Metamaterials have progressed from academic curiosities\cite{Shalaev,Padilla200628} to candidates for real-world applications.\cite{0268-1242-20-7-018, 2003OExpr..11.2549K} Emerging metamaterial applications range from radio frequency (RF)\cite{PSSB:PSSB200674511} communications to millimeter radar.\cite{sato2008}  One key technique which promises to further the applicability of metamaterials is tunability.  Operational frequency tuning has been presented as one solution to the narrow bandwidth often present in metamaterial devices.\cite{Schurig10112006,Liu23032007}  Frequency-agile metamaterial designs have been demonstrated across a wide spectrum from microwave \cite{Shadrivov:06} to near-visible frequencies.\cite{Dicken:09} 

To date, tuning has generally been homogeneously implemented across the entire device as a whole.  Developing techniques for controllable spatially variable tuning will present the possibility of devices with a reconfigurable Gradient Index of Refraction (GRIN).  GRIN devices have already proven an attractive area for metamaterials,\cite{driscoll:081101} as the metamaterial design process naturally allows for the control needed to fabricate GRIN structures.  Additionally, use of spatially nonuniform tuning can leverage the narrow bandwidth of metamaterials in a unique way.  For narrow-band operation, minor adjustments in the resonance frequency of a metamaterial can translate to large changes in the index of refraction at that frequency.  Overall, spatial control of resonance tuning allows for post-fabrication modification of the index of refraction and therefore the creation of a reconfigurable gradient. 

\begin{figure}
\includegraphics{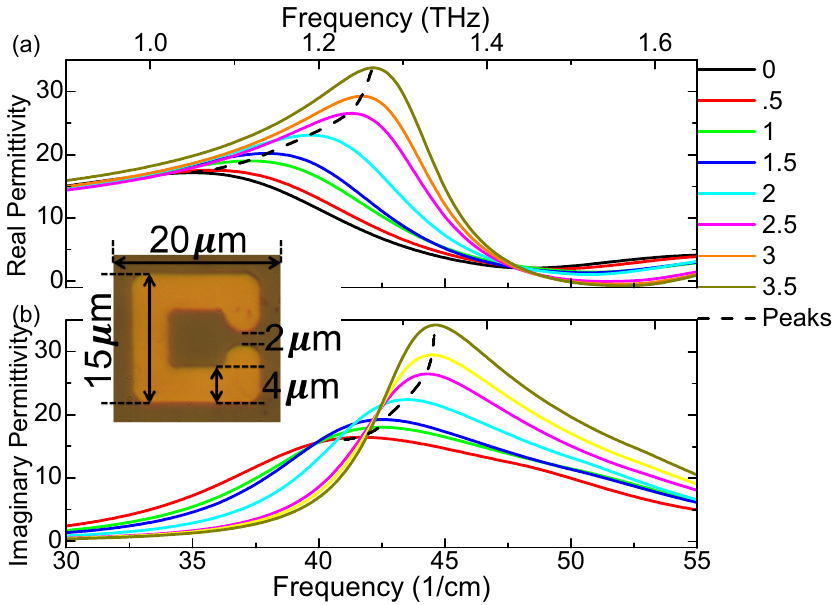}
\caption{(Color online) Extracted values for real (a) and imaginary (b) permittivity. Position of maximum permittivity indicated by dashed lines. (Inset) Schematic of SRR unit cell.\label{fig:permittivity}}
\end{figure}

In this work, we demonstrate a spatially reconfigurable THz hybrid metamaterial with vanadium dioxide (VO$_2$) and split ring resonators (SRRs) as constituent elements. The SRR has been the ``fruit fly'' of metamaterials research, allowing for convenient implementation of optical characteristics which are unattainable without metamaterials.\cite{Yen05032004,Shalaev,PhysRevLett.84.4184} Our device is composed of an array of 100 nm thick gold SRRs (dimensions shown in figure \ref{fig:permittivity}) lithographically fabricated on 90 nm thick VO$_2$ grown on a sapphire substrate. VO$_2$ undergoes an insulator to metal transition\cite{RevModPhys.83.471} which can be triggered thermally electrically\cite{Kim2010jap} or optically.\cite{PhysRevB.79.075107} The phase transition is hysteretic, and therefore, changes in the conductivity of VO$_2$ generally persist, provided the device temperature is maintained. Hybrid metamaterial-VO$_2$ devices benefit from this memory,\cite{Driscoll18092009} and from the large tuning dynamic range achievable with VO$_2$.  Persistent tuning uses this memory to eliminate the need for continuous stimulation or repeated excitation of the device.  Our measurements demonstrate the ability to perform persistent tuning of a hybrid SRR-VO$_2$ device with a spatial configuration, writing a gradient in the permittivity of the hybrid metamaterial.

To interrogate our metamaterial, we performed infrared transmission spectroscopy using a home-built broadband IR microscope.  Linearly polarized light from a mercury lamp was focused via reflecting optics to a normally-incident spot $\sim$100$\mu$m in diameter. In far-infrared, this spot facilitates nearly diffraction-limited measurements of the metamaterial local response.  The sample was mounted on a translator allowing for linear movement across the entire device.  Integration of a temperature stage with proportional-integral-derivative controller (PID) feedback allowed us to control device temperature to better than $\pm$0.05 K.  For these experiments, we kept the device at a constant temperature of 339.5 K (just below the temperature of the insulator-metal transition in VO$_2$) to allow for maximal transition during the voltage application discussed below.  

Prior to the application of any current pulse, calibration transmission spectra were collected at positions across the sample at 339.5 K. In these spectra, the peak position of 1-\textit{T}($\omega)$ (transmission) indicates the resonance frequency of the hybrid metamaterial. In our setup, the device was oriented to electrically excite the lowest order RLC-oscillation mode of the SRR.  The increasing permittivity of VO$_2$ as its phase transition progresses increases the capacitance of each SRR. This in turn reduces the effective time constant of the RLC-oscillation mode thereby decreasing the resonance frequency of the hybrid SRR-VO$_2$ device. Additionally, losses in the metallic regions of VO$_2$ cause damping of the resonance. These calibration spectra are nearly identical with uniform resonance positions, indicating the homogeneous (on a mesoscopic effective medium scale) initial state of our device.

\begin{figure}
\includegraphics{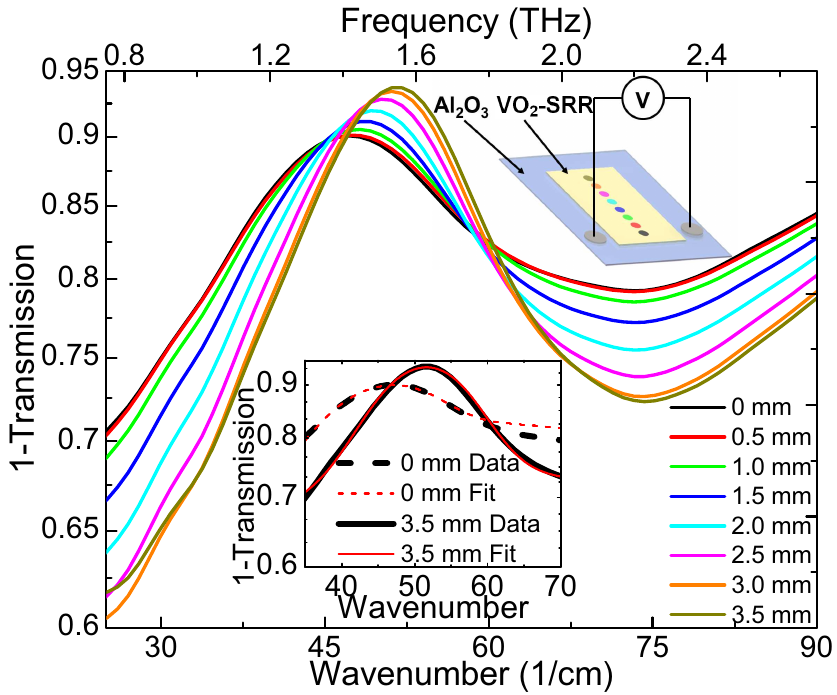}
\caption{(Color online) Transmission measurements taken along the hybrid metamaterial at 339.5 K. 0 mm indicates a position midway between the contacts. (Bottom inset) Measured transmission curves along with fits. (Top inset) Schematic of device indicating contact geometry and measurement locations. \label{fig:transmission} }
\end{figure}

To electrically write a spatial gradient to the device, we applied a square voltage pulse of 175 V $\times$ 2.5 s across the contact points as shown in Figure \ref{fig:transmission} (top inset). This voltage pulse did not significantly affect \textit{average} temperature of the device as the total input energy is small compared to the thermal inertia of the sapphire substrate.  However, as current flowed almost exclusively through the VO$_2$-SRR layer, local transient heating of this layer occurred.  While the dominant switching mechanism in this device is thermal, there may be an additional contribution due to electrical switching from the applied voltage. Inquiry into the microscopic mechanisms of switching is an active area of VO$_2$ research\cite{PhysRevB.79.075107,Kim2010jap} but is beyond the scope of our work, which focuses on the utility and applications of switching effects. As VO$_2$ undergoes its transition from insulator to metal,\cite{1367-2630-6-1-052} we expect the peak of the resonance to redshift and become damped. \cite{4838777} Such transient heating leaves a persistent change due to the hysteresis in VO$_2$.\cite{Driscoll18092009, 4838777}

The placement of electrical contacts was designed to induce a gradient in the current density passing through the VO$_2$-SRR layer.  Using our translation stage, we probed 8 locations on the sample, ranging from directly between the contacts (0 mm) to +3.5 mm above the contact path as shown in Figure \ref{fig:transmission} (top inset).  In Figure \ref{fig:transmission}, we plot transmission spectra at these 8 positions. The redshift of the absorption resonance is more prominent for positions in close proximity to the contacts. This is consistent with the idea that the (percolative) phase transition in VO$_2$ has progressed further near the contacts. The maximum resonance redshift is about 10\% of the center frequency, and nearly 50\% of that obtained from the entire insulator-metal transition.\cite{4838777} Through a comparison of the measured resonance frequencies to those in a previous work using the same device,\cite{4838777} we estimated an effective temperature gradient,  during and shortly after the current pulse, ranging from $\sim$342 K at 0 mm, to $\sim$340 K at 3.5 mm. Additionally, this experiment demonstrates that metamaterials can be probed and meaningfully modified on spatial scales on the order of the wavelength.  The total spatial extent of our observed gradient is 3.5 mm, with 8 data points over this distance.  Thus, these distinct spectra are recorded with a spacing of just under 1.5$\lambda$ ($\sim300\mu$m at 1THz).

To better quantify the potential of such gradients, we retrieved the optical parameters from the transmission curves by fitting.  Permittivity gives a better visualization of the gradient in material response than transmission, especially in a single-layer metamaterial where overall transmission is generally high.  The retrieval procedure is a two-oscillator fitting using Lorentzian oscillators \boldmath{$\bar{\epsilon}$} and \boldmath{$\bar{\mu}$} weighted by spatial-dispersion cofactors \cite{PhysRevE.76.026606,PhysRevE.81.036605} to give a close approximation to the homogenized effective permittivity, \boldmath{$\epsilon_{\text{eff}}$}, and permeability, \boldmath{$\mu_{\text{eff}}$} of the metamaterial layer:
\begin{equation}
\label{eq:1}
\epsilon_{\text{eff}} = \bar{\epsilon}\frac{(\theta/2)}{\sin(\theta/2)}\left[\cos(\theta/2)\right]^{-S_b}
\end{equation}
\begin{equation}
\label{eq:2}
\mu_{\text{eff}} = \bar{\mu}\frac{(\theta/2)}{\sin(\theta/2)}\left[\cos(\theta/2)\right]^{S_b}
\end{equation}
where $\theta$ is the phase advance across one cell, and $S_b$=1 for electric resonators or $S_b$=-1 for magnetic resonators. The resonance frequency, strength, and damping constant of two oscillators\cite{0953-8984-14-25-307} underlying the effective optical parameters were modified to obtain good agreement between measured and fitted spectra. While better agreement may have been obtained through the use of a large number of oscillators, this procedure increases numerical complexity, and can often obscure insight into the underlying physical processes involved in the material response, here the magnetic ($\omega_0$) and electric ($\omega_1$) modes of the SRR.\cite{PhysRevLett.96.107401} These oscillator parameters were then used to determine permittivity and permeability of our device.  Resultant fits for the most shifted (0 mm) and least shifted (3.5 mm) transmission lines are shown in Figure \ref{fig:transmission} (bottom inset). The fits capture the most prominent features of the transmission spectra. They also model the pulse-induced redshift and damping of the resonance. 

\begin{figure}
\includegraphics{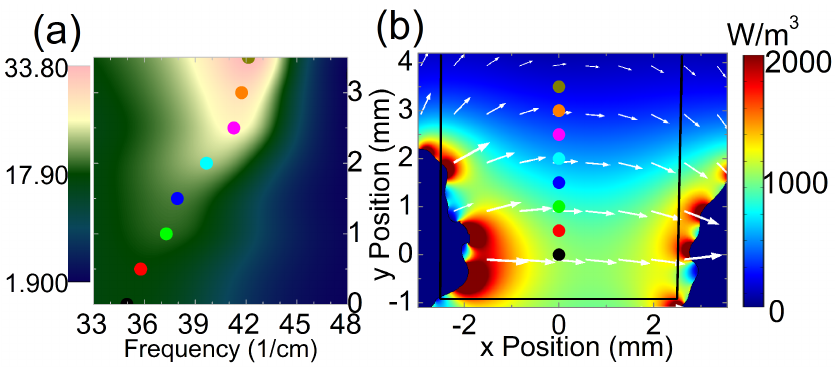}
\caption{(Color online) a) Spatial map of real permittivity varying with vertical position and frequency. Resonance frequency locations indicated by colored points. b) Simulation of current flow through device. Color indicates power density (W/m$^3$) and current flow is shown in the vector field (A/m$^2$). Black rectangular outline indicates the location of the SRR array and VO$_2$. Contact locations and shapes indicated by outlines at the bottom of the device. \label{fig:current} }
\end{figure}

The real and imaginary permittivity extracted from these fits are shown in Figure \ref{fig:permittivity}.  These figures illustrate the range of complex permittivity values simultaneously present at different locations in the device.  By interpolating between the 8 points where the spectra were obtained, we extracted a spatial map of the metamaterial permittivity response, shown in Figure \ref{fig:current} (a).  At locations nearer to the contacts, the resonance becomes increasingly redshifted due to the increasing number and density of metallic puddles in VO$_2$.  The overall frequency tuning range of our device compares favorably with many techniques for frequency-agile metamaterials.\cite{Chen:2010p35,4838777} At the initial resonance frequency ($\sim$1.23 THz), the overall percent change in real permittivity is 70\% across the sample, enough to satisfy many potential GRIN device applications.  

The shift in resonance position in our transmission spectra is a result of the application of a transient voltage pulse as described previously. Current flowing through the device heats the VO$_2$ with power dissipated given as $P=I^2 R$. We expect that areas of the device nearer to the contacts were exposed to a larger current density, and as a result the VO$_2$ will have progressed through a greater amount of its phase transition. To model this situation, we performed a simulation of current flow and power dissipation in our device using COMSOL, a finite element simulation program. The results of this simulation are shown in Figure \ref{fig:current} (b).  As expected, the greatest power dissipation occurs in areas near the contacts. These results fit with our picture of a spatial gradient of permittivity induced by inhomogeneous excitation of the VO$_2$ phase transition. 

Our SRR-VO$_2$ device is the simplest hybrid metamaterial structure with which to explore advantages of spatial gradient tuning. With it, we have shown the ability to induce and spectroscopically probe a persistent spatial gradient within a previously uniform device. Use of finer, or even pixel-by-pixel, control over switching would allow for creation of more precise gradients. Such pixel-level switching nearly necessitates the existence of memory within the device as the difficulty of repeated or sustained excitation would make detailed switching unfeasible. Future flexibility in the optical parameters of metamaterials as a result of tuning occurring on the pixel level will enable the creation of dynamic hybrid metamaterial structures suitable for a wide range of applications.
\newline

The authors wish to acknowledge funding for this project from AFOSR and the creative research project of ETRI.

\end{document}